\documentclass[12pt]{iopart}
\usepackage{iopams}
\usepackage{setstack}
\usepackage[centertableaux]{ytableau}
\usepackage[vcentermath]{youngtab}
\usepackage{graphicx}
\usepackage{multirow}
\usepackage{hyperref}
\usepackage{longtable}
\usepackage{float}
\usepackage{makecell}
\usepackage{braket}
\usepackage{subfigure}
\usepackage{cite}
\begin{document}

\title{A possible interpretation of $\Lambda$ baryon spectrum with pentaquark components}
%\author{Authors}

%\address{IOP Publishing, Temple Circus, Temple Way, Bristol BS1 6HG, UK}
%\ead{submissions@iop.org}

\author{Kai Xu$^{1}$, Attaphon Kaewsnod$^{1}$, Zheng Zhao$^{1}$, Thiri Yadanar Htun$^{1,2}$, Ayut Limphirat$^1$, Yupeng Yan$^1$}

\address{$^1$ School of Physics and Center of Excellence in High Energy Physics and Astrophysics, Suranaree University of Technology, Nakhon Ratchasima 30000, Thailand}
\address{$^2$ Department of Physics, University of Mandalay, 05032 Mandalay, Myanmar}
\ead{kaix@g.sut.ac.th, ayut@g.sut.ac.th and yupeng@g.sut.ac.th}

\vspace{10pt}
\begin{indented}
\item[]October 2022
\end{indented}

\begin{abstract}
The $\Lambda$ baryon spectrum is studied within the SU(3) flavor symmetry in a constituent quark model. We found that it is rather difficult to accommodate some negative-parity $\Lambda$ resonances as single $q^2s$ ($q = u,\,d$ quarks) states in the conventional three-quark picture. The ground $q^3s\bar q$ pentaquark mass spectrum is evaluated and a possible interpretation is proposed in the work: the observed $\Lambda(1405)1/2^{-}$, $\Lambda(1670)1/2^{-}$ and $\Lambda(1800)1/2^{-}$ are three-state mixtures of two $p$-wave $q^2s$ states and one ground $q^3s\bar q$ pentaquark state, so are the $\Lambda(1520)3/2^{-}$, $\Lambda(1690)3/2^{-}$ and $\Lambda(2050)3/2^{-}$ resonances.
%The higher partner states from the three-state mixtures with the lowest $q^3s\bar q$ pentaquark states are at the energy 1828 MeV for spin $1/2^{-}$ and 1977 MeV for spin $3/2^{-}$ which may correspond to the observed  and  separately.

%While $\Lambda(2080)5/2^{-}$, $\Lambda(2100)7/2^{-}$ and $\Lambda(2325)3/2^{-}$ would be l=3 three-quark $\Lambda$ states,
%\indent We study both the $\Lambda$ baryon spectrum and $q^3s\bar q$ (q = u,d quark) pentaquark states in SU(3) flavor symmetry with a general group theory approach in the constituent quark model. The $\Lambda$ baryon spectrum is newly reproduced in the normal three-quark picture while some negative-parity excited $\Lambda$ resonances are studied as the mixtures of two three-quark ($q^2s+q^2s$) states, one three-quark and one pentaquark ($q^2s+q^3s\bar q$) state, and two three-quark and one pentaquark ($2q^2s+q^3s\bar q$) states.   By diagonalizing the mass matrices of possible mixing states, we provide a possible interpretation that the $\Lambda(1520)3/2^{-}$ and $\Lambda(1690)3/2^{-}$ are the two three-quark negative-parity mixing states, and the $\Lambda(1405)1/2^{-}$, $\Lambda(1670)1/2^{-}$ and $\Lambda(2000)1/2^{-}$ are the three-body mixtures including two three-quark and one ground $q^3s\bar q$ pentaquark states.
%(Based on the assumption that the bare and physical $\Lambda$ baryon states are related by the unitary transformations,)

\end{abstract}

\section{Introduction}\label{sec:Int}

\indent In recent decades, new experimental approaches have been applied to study the $\Lambda$ resonances. Experimental data are mainly from $K^- p$ invariant mass spectra of hyperons below 1.7 GeV \cite{Star2001,Prakhov2004,Prakhov2009}, partial wave analyses of $K^- p$ reactions with energies from 1480 to 2100 MeV \cite{Zhang12013} and photoproduction of $\Lambda(1405)$ from JLab \cite{Moriya2013,Moriya2014}. The $\Lambda$ resonances are listed in table \ref{tab:lambdaR} \cite{PDG2020}.

Recent data analyses have confirmed most 3* and 4* $\Lambda$ states, but derived rather different information of some little known states. Recent couple channel analyses from the Kent group (KSU) and JLab have extracted more hyperons \cite{Zhang22013,Fern2016,Kama2014,Kama2015,Matveev2019,Sara2019,Aniso2020,Klempt2020}. The Kent group found some new states like  $\Lambda$(1710)$1/2^+$ and $\Lambda$(2050)$3/2^-$, which were listed in Particle Data Group (PDG)~\cite{PDG2020} with one star, by fitting partial-wave amplitudes using multichannel parametrization \cite{Zhang22013}. A similar couple-channel fit has been done by JPAC group \cite{Fern2016}, where $\Lambda$(2050) is confirmed by this analysis but $\Lambda$(1800) is not seen. A dynamical coupled-channel model applied by the Osaka group \cite{Kama2014,Kama2015} results in no clear existence of $\Lambda$(1710), $\Lambda$(1800), $\Lambda$(2000) and $\Lambda$(2050). Bonn-Gatchina (BnGa) group has recently reported $\Lambda$(1710)$1/2^+$, $\Lambda$(2070)$3/2^+$ and $\Lambda$(2080)$5/2^-$, suggesting all the states together with $\Lambda(1405)$$1/2^-$ as SU(3) singlet states \cite{Matveev2019,Sara2019,Aniso2020,Klempt2020}. The experimental data from $K^- p$ induced reactions have made a great contribution to our knowledge of the $\Lambda$ baryon spectrum, but it is still insufficient to reveal the nature of all $\Lambda$ resonances.

\begin{table}[b]
\caption{$\Lambda$ resonances (Data from PDG).}
\begin{center}\label{tab:lambdaR}
\vspace*{-0.15cm}
\begin{indented}
	\item[]
\begin{tabular}{lcccc}
\br
 Particle  & Status & $J^{P}$ &  $M_{BW}^{exp}$ ${\rm MeV}$ & $\Gamma_{BW}^{exp}$ ${\rm MeV}$
\\
\hline
 $\Lambda(1116)$ & **** & $\frac{1}{2}^{+}$ & 1116 & -
\\
 $\Lambda(1600)$ & **** & $\frac{1}{2}^{+}$ & 1570-1630 & 150-250
\\
 $\Lambda(1710)$ & *  &$\frac{1}{2}^{+}$ & 1700-1726 & 138-222
\\
 $\Lambda(1810)$ & *** &$\frac{1}{2}^{+}$ & 1740-1840 & 50-170
\\
 $\Lambda(1820)$ & **** &$\frac{5}{2}^{+}$ & 1815-1825 & 70-90
\\
 $\Lambda(1890)$ & ****  &$\frac{3}{2}^{+}$ & 1870-1910 & 80-160
\\
 $\Lambda(2070)$ & * &$\frac{3}{2}^{+}$ & 2046-2094 & 320-420
\\
 $\Lambda(2085)$ & **  &$\frac{7}{2}^{+}$ & 2000-2040 & 130-190
\\
 $\Lambda(2110)$ & ***  &$\frac{5}{2}^{+}$ & 2050-2130 & 200-300
\\
 $\Lambda(2350)$ & ***  &$\frac{9}{2}^{+}$ & 2340-2370 & 100-250
\\
\mr
 $\Lambda(1380)$ & ** &$\frac{1}{2}^{-}$ & 1310-1340 & -
\\
 $\Lambda(1405)$ & **** &$\frac{1}{2}^{-}$ & 1404-1406 & 48.5-52.5
\\
 $\Lambda(1520)$ & **** &$\frac{3}{2}^{-}$ & 1518-1520 & 15-17
\\
 $\Lambda(1670)$ & **** &$\frac{1}{2}^{-}$ & 1670-1678 & 25-35
\\
 $\Lambda(1690)$ & **** &$\frac{3}{2}^{-}$ & 1685-1695 & 60-80
\\
 $\Lambda(1800)$ & *** &$\frac{1}{2}^{-}$ & 1750-1850 &150-250
\\
 $\Lambda(1830)$ & **** &$\frac{5}{2}^{-}$ & 1820-1830 & 60-120
\\
 $\Lambda(2000)$ & * &$\frac{1}{2}^{-}$ & 2000-2060 & 100-150
\\
 $\Lambda(2050)$ & * &$\frac{3}{2}^{-}$ & 2034-2078 & 432-554
\\
 $\Lambda(2080)$ & * &$\frac{5}{2}^{-}$ & 2069-2095 & 152-210
\\
 $\Lambda(2100)$ & **** &$\frac{7}{2}^{-}$ & 2090-2110 & 100-250
\\
 $\Lambda(2325)$ & * &$\frac{3}{2}^{-}$ & 2307-2347 & 120-200
\\
\br
\end{tabular}
\end{indented}
\end{center}
\end{table}

The $\Lambda$ spectra have been studied in various quark models theoretically, for example, the Isgur-Karl model \cite{Isgur1978PRD,Isgur1979PRD,Grome1983} which includes the one-gluon exchange quark-quark interaction, models taking into account the Goldstone boson exchange interaction \cite{Gloz1998}, models considering the relativistic instanton-induced quark-quark interaction \cite{Loring2001}, and relativistic quark models with quark-diquark interactions \cite{Santo2015,Fau2015}.

The traditional constituent quark models always have difficulties in describing the low mass of $\Lambda(1405)$ \cite{Capstick2000} in the $q^3$ picture since a heavier strange quark is contained. The $\Lambda$(1405) resonance, since discovered in the $\pi \Sigma$ invariant mass spectrum \cite{Alston1961} in 1960s, has been an interesting state in the hyperon spectrum, the nature of the resonance is still under debate. $\Lambda$(1405) was first generated dynamically as a $N\bar K$  quasi-bound molecular state \cite{Dalitz1960,Dalitz1967}, then followed a large number of studies in chiral unitary approach (ChUA)  \cite{Oller2001,Oller2003,Roca2013,Oset2013,Mai2015} and other chiral dynamics on this resonance \cite{kaiser1997,Oset2002,Cieply2010,Cieply2012,Cieply2016,Miyahara2018,Ren2021,Wang2021,Ikeda2011,Ikeda2011two,Guo2013,Mai2013,Bruns2022}. By solving the coupled channel Lippmann-Schwinger equation with a chiral meson-baryon Lagrangian at the next-to-leading order, the $\Lambda$(1405) resonance as a quasi-bound state of $N\bar K$ described the low-energy $K^-p$ scattering data well \cite{kaiser1997}. Limited to the lowest-order chiral Lagrangian with coupled channel method, $\Lambda$(1405) was identified as quasi-bound states of $N\bar K$ using the N/D method and dispersion relations \cite{Oset2002}. The two-poles structure of $\Lambda(1405)1/2^-$ was firstly proposed in the complex energy plane for the different Riemann sheets using the ChUA in 2001 \cite{Oller2001}. Then a variety of investigations on the two-poles structure of the $\Lambda$(1405) resonance were carried out. The contributions of the two-pole structure to the lineshape of $\Lambda$(1405) were shown in Refs. \cite{Oller2003,Cieply2010}. After the SIDDHARTA kaonic hydrogen result \cite{Bazzi2011,Bazzi2012} was reported, different works \cite{Cieply2012,Ikeda2011,Ikeda2011two,Guo2013,Mai2013,Miyahara2018} revealed the real and imaginary parts of the two poles of $\Lambda(1405)1/2^-$. New fits including the CLAS photoproduction data were carried out in the works \cite{Roca2013,Oset2013,Mai2015,Cieply2016} and the origin of two poles of $\Lambda(1405)1/2^-$ were discussed as well. In recent years, the first high-mass pole of $\Lambda(1405)1/2^-$ was shown to be mainly a $\bar K N$ molecular, the second pole was interpreted as a composite state of mainly $\pi \Sigma$ with tiny component $\bar K N$ in Refs. \cite{Ren2021,Wang2021,Bruns2022}, and the influence of SU(3) flavor symmetry was analyzed in the last work \cite{Bruns2022}. Exotic configurations for $\Lambda(1405)$ have also been proposed, such as a mixed pentaquark and three-quark state \cite{Nakamura2008}, a hybrid baryon \cite{Azi2018}, a mixed strange hybrid/three-quark baryon \cite{Kiss2011} which all applied the method of the QCD sum rule. A mixed three-quark and meson-baryon molecular states for the $\Lambda(1405)$ resonance are studied in Refs. \cite{Arima1994,Sachi2007} in quark models. The electromagnetic form factors of the $\Lambda(1405)$ resonance were calculated in Lattice QCD for the first time to reveal the inner structure of $\Lambda(1405)$ \cite{Hall2015,Hall2017}.

In the present work we extend the non-relativistic quark model employed in the works \cite{Kai2019PRC,Kai2020PRD} to study the $\Lambda$ mass spectra including light $q^3s\bar q$ pentaquark components. The lowest two pairs of negative-parity $\Lambda$ resonances are reproduced by mixing up the three-quark $\Lambda_1^*$ (flavor-singlet), $\Lambda_8$ (flavor-octet) states with the ground $q^3s\bar q$ pentaquark states.
The paper is organized as follows. In section \ref{sec:GTM} we briefly introduce the Hamiltonian for multi-quark systems including the spin-orbit couplings. The masses of low-lying $q^2s$ states are also calculated in section \ref{sec:GTM}. In section \ref{sec:LPM}, we derive the mass spectra of ground $q^3s \bar q$ pentaquark states and propose an interpretation of the lowest two pairs of negative-parity $\Lambda$ resonances by introducing the ground $q^3s \bar q$ pentaquark components in the three-state mixtures. A discussion and a short summary are given in sec.~\ref{sec:SUM}.
%with quantum numbers, N=2n+1, L=1 and L=3 in Harmonic oscillator bases
%The mixtures of two p wave three-quark $q^2s$ states and one ground $q^3s\bar q$ pentaquark states are discussed.

\section{\label{sec:GTM} $\Lambda$ resonances as $q^2s$ states}

\indent We calculate the mass of low-lying $q^2s$ states in the Hamiltonian, which is an extend of the model in Ref. \cite{Kai2019PRC} by including the spin-orbit interactions,
\begin{eqnarray}\label{eqn::ham}
&H =H_0+ H_{hyp}^{OGE}+H^{SO}, \nonumber \\
&H_{0} =\sum_{k=1}^{N} (m_k+\frac{p_k^2}{2m_{k}})+\sum_{i<j}^{N}(-\frac{3}{8}\lambda^{C}_{i}\cdot\lambda^{C}_{j})(A_{ij} r_{ij}-\frac{B_{ij}}{r_{ij}}),  \nonumber \\
&H_{hyp}^{OGE} = -C_{OGE}\sum_{i<j}\frac{\lambda^{C}_{i}\cdot\lambda^{C}_{j}}{m_{i}m_{j}}\,\vec\sigma_{i}\cdot\vec\sigma_{j}, \nonumber \\
&H^{SO}= 2k_i\,d_j\;\vec L \cdot \vec S.
%&H_{q^3}^{SO}= \sum_{i<j}\frac{1}{2r_{ij}}\frac{d(V-S/3)}{d r_{ij}}(\frac{1}{m^2_i}(\vec {S_i} \cdot \vec {r_{ij}} \times \vec{p_i})- \nonumber \\
%& \frac{1}{m^2_j}(\vec {S_j} \cdot \vec {r_{ij}} \times \vec{p_j}) -\frac{2}{m_i m_j}[(\vec {S_i} \cdot \vec {r_{ij}} \times \vec{p_j})-(\vec {S_j} \cdot \vec {r_{ij}} \times \vec{p_i})] ).
\end{eqnarray}
where $A_{ij}$ and $B_{ij}$ are mass-dependent coupling parameters, taking the form
\begin{eqnarray}
A_{ij}= a \sqrt{\frac{m_{ij}}{m_u}},\;\;B_{ij}=b \sqrt{\frac{m_u}{m_{ij}}}.
\end{eqnarray}
with $m_{ij}$ being the reduced mass of $i$th and $j$th quarks, defined as $m_{ij}=\frac{m_i m_j}{m_i+m_j}$.
The hyperfine interaction, $H_{hyp}^{OGE}$ includes only one-gluon exchange contribution (OGE),
where $C_{OGE} = C_m\,m_u^2$, with $m_u$ being the constituent $u$ quark mass and $C_m$ a constant. $\lambda^C_{i}$ in the above equations are the generators of color SU(3) group. The three model coupling constants and two constituent quark masses are taken from the previous work \cite{Kai2020PRD},
\begin{eqnarray}\label{eq:nmo}
&
m_u = m_d = 327 \ {\rm MeV}\,, \quad
m_s = 498 \ {\rm MeV}\,, \nonumber\\
&
C_m   =  18.3 \ {\rm MeV}, \quad
a     = 49500 \ {\rm MeV^2}, \quad
b     =  0.75 \nonumber\\
\end{eqnarray}

$H^{SO}$ in equation (\ref{eqn::ham}) is the
spin-orbit interaction, taking the simplified form as in Refs. \cite{Isgur1978PRD,Isgur1979PRD,Grome1983}, where $L$ and $S$ are the total orbital angular and spin operators of baryon states.
$d_j$ are $SU(3)_{F}$ flavor multiplet-dependent constants, taking the values,
\begin{eqnarray}\label{eq:slm1}
d_1 = 2/3 \,, \quad
d_8 = 1 \,, \quad
d_{10} = 1/3
\end{eqnarray}
$k_i$ are $SU(6)_{SF}$ spin-flavor supermultiplet-dependent constants, which are determined by fitting the theoretical results to the mass splitting of spin-orbit pairs in the
$\Lambda$, $\Sigma$ and $\Sigma^{*}$ spectra in the pure $q^3$ picture. 
The mass difference between the two $\Lambda(70,{}^4 8,1,1^-)$ states $\frac{1}{2}^{-}$ $\Lambda$(1800) and $\frac{5}{2}^{-}$ $\Lambda$(1830) leads to $k_1 \approx 4 $ MeV while the gap between the $\frac{3}{2}^{-}$ $\Sigma$(1580) and $\frac{1}{2}^{-}$ $\Sigma$(1620) states results in $k_1 \approx 3 $ MeV. The $(70, 1^{-})$ and $(20, 1^{+})$ supermultiplets have a direct relation in the lowest harmonic oscillation approximation, which results in $k_2\approx 1.5\; k_1\approx 6\; {\rm MeV}$. $k_3$ is determined about $-14$ MeV by the $\Lambda(56,{}^2 8,2,2^+)$ states $\Lambda(1820)$ and $\Lambda(1890)$, and about $-5 $ MeV by the $\Sigma(56,{}^2 8,2,2^+)$ states $\Sigma(1915)$ and $\Sigma(1940)$.
The mass gap between the $\Lambda(70,{}^4 8,2,2^+)$ states $\Lambda(2110)$ and $\Lambda(2085)$ leads to $k_4 \approx -14 $ MeV, and the one between the $\Sigma(70,{}^4 8,2,2^+)$ states $\Sigma(2070)$ and $\Sigma(2030)$ leads to $k_4 \approx -6 $ MeV. 
In this work, we use averaged values,  
\begin{eqnarray}\label{eq:slm2}
& k_1 ([70, 1^{-}]) = 4 \ {\rm MeV}\,,
k_2 ([20, 1^{+}]) = 6 \ {\rm MeV}\,, \nonumber\\
&
k_3 ([56, 2^{+}])  = -10 \ {\rm MeV}\,,
k_4 ([70, 2^{+}])) = -10 \ {\rm MeV}
\end{eqnarray}

The positive-parity and negative-parity $\Lambda$ resonances of the first and second excitation bands are studied in the Hamiltonian in equation (\ref{eqn::ham}) in the three-quark picture. The color-orbital-spin-flavor wave functions of $q^2s$ baryon states in the $SU(6)_{SF}$ representations, which are applied in the study, are listed in Appendix \ref{ap:Lams}. The theoretical masses and tentative matchings between data and the theoretical results are shown in table \ref{tab:posN}.

\begin{table}[t]
\caption{Theoretical $\Lambda$ resonances matched with data up to the second excitation band. $\Lambda^*$ stands for the flavor-singlet states. $\Gamma$ and $D$ stands respectively for the $SU(6)_{SF}$ and $SU(3)_{F}$ multiplets, $N$, $S$, $L^P$ and $J^P$ are the principle quantum number, total spin, orbital and angular momentum with $P$ being the parity.}
\label{tab:posN}
\begin{center}
\begin{indented}
	\item[]
\begin{tabular}{lcccc}
\br
 $(\Gamma, {}^{2S+1} D,N,L^P)$  & Status & $J^{P}$ &  $M^{exp}  {\rm (MeV)}$ & $M^{cal} {\rm (MeV)}$
\\
\hline
 $\Lambda(56,{}^2 8,0,0^+)$ & **** & $\frac{1}{2}^{+}$ & $\Lambda(1116)$ & 1112
\\
 $\Lambda(56,{}^2 8,2,0^+)$ & **** & $\frac{1}{2}^{+}$ & $\Lambda(1600)$ & 1689
\\
 $\Lambda(56,{}^2 8,2,2^+)$ & **** &$\frac{5}{2}^{+}$ & $\Lambda(1820)$ & 1825
\\
 $\Lambda(56,{}^2 8,2,2^+)$ & **** & $\frac{3}{2}^{+}$ & $\Lambda(1890)$ & 1875
\\
 $\Lambda(20,{}^2 8,2,1^+)$ & - &$\frac{1}{2}^{+}$ & missing & 1927
\\
 $\Lambda(20,{}^4 8,2,1^+)$ & - &$\frac{3}{2}^{+}$ & missing & 1945
\\
 $\Lambda^*(20,{}^4 1,2,1^+)$ & - &$\frac{1}{2}^{+}$ & missing & 2162
\\
 $\Lambda^*(20,{}^4 1,2,1^+)$ & - &$\frac{3}{2}^{+}$ & missing & 2174
\\
 $\Lambda^*(20,{}^4 1,2,1^+)$ & - &$\frac{5}{2}^{+}$ & missing & 2194
\\
 $\Lambda(70,{}^2 8,2,0^+)$ & *** &$\frac{1}{2}^{+}$ & $\Lambda(1810)$ & 1821
\\
 $\Lambda(70,{}^4 8,2,0^+)$ & * &$\frac{3}{2}^{+}$ & $\Lambda(2070)$ & 2081
\\
 $\Lambda^*(70,{}^2 1,2,0^+)$ & - &$\frac{1}{2}^{+}$ & $\Lambda(1710)$ & 1838
\\
 $\Lambda(70,{}^2 8,2,2^+)$ & **** &$\frac{3}{2}^{+}$ & $?\Lambda(1890)$ & 1922
\\
 $\Lambda(70,{}^2 8,2,2^+)$ & **** &$\frac{5}{2}^{+}$ & $?\Lambda(1820)$ & 1872
\\
 $\Lambda(70,{}^4 8,2,2^+)$ & -  &$\frac{1}{2}^{+}$ & missing & 2242
\\
 $\Lambda(70,{}^4 8,2,2^+)$ & -  &$\frac{3}{2}^{+}$ & missing & 2212
\\
 $\Lambda(70,{}^4 8,2,2^+)$ & ***  &$\frac{5}{2}^{+}$ & $\Lambda(2110)$ & 2142
\\
 $\Lambda(70,{}^4 8,2,2^+)$ & **  &$\frac{7}{2}^{+}$ & $\Lambda(2085)$ & 2092
\\
 $\Lambda^*(70,{}^2 1,2,2^+)$ & - &$\frac{3}{2}^{+}$ & missing & 1929
\\
 $\Lambda^*(70,{}^4 1,2,2^+)$ & - &$\frac{5}{2}^{+}$ & missing & 1896
\\
\mr
 $\Lambda(70,{}^2 8,1,1^-)$ & $\phantom{-}$ &$\frac{1}{2}^{-}$ & $\phantom{-}$ &  1579
 \\
 $\Lambda(70,{}^2 8,1,1^-)$ & $\phantom{-}$ &$\frac{3}{2}^{-}$ & $\phantom{-}$ &  1591
\\
 $\Lambda(70,{}^4 8,1,1^-)$ & *** &$\frac{1}{2}^{-}$ & $\Lambda$(1800) &  1810
\\
 $\Lambda(70,{}^4 8,1,1^-)$ & **** &$\frac{5}{2}^{-}$ & $\Lambda$(1830) &  1842
\\
 $\Lambda(70,{}^4 8,1,1^-)$ & - &$\frac{3}{2}^{-}$ & missing & 1822
\\
 $\Lambda^*(70,{}^2 1,1,1^-)$ & $\phantom{-}$ &$\frac{1}{2}^{-}$ & $\phantom{-}$ &  1582
 \\
 $\Lambda^*(70,{}^2 1,1,1^-)$ & $\phantom{-}$ &$\frac{3}{2}^{-}$ & $\phantom{-}$ &  1590
\\
\br
\end{tabular}
\end{indented}
\end{center}
\end{table}

In the second excitation band shown in table \ref{tab:posN}, the positive-parity $\Lambda$ resonances belonging to the 20-supermultiplet for both the singlet and octet flavors are missing. $\Lambda$(1820) and $\Lambda$(1890) are matched with the spin $\frac{5}{2}$ and $\frac{3}{2}$ states in the 56-multiplets respectively, but it can not be ruled out that the two resonances are matched with the 70-multiplet $\frac{5}{2}$ and $\frac{3}{2}$ states. The 3-star $\Lambda$(1810) state is assigned as a $\Lambda(70,{}^2 8,2,0^+)$ state in this work. This resonance $\Lambda$(1710) reported by KSU \cite{Zhang22013} in 2013 is interpreted as the $\Lambda^*(70,{}^2 1,2,0^+)$ singlet state in Ref. \cite{Klempt2020}, but it can't be described by the relativistic interacting quark-diquark models \cite{Santo2015,Fau2015}. Though a higher mass is predicted for the flavor-singlet state $\Lambda^*(70,{}^2 1,2,0^+)$ in the work, we may tentatively assign the 1-star $\Lambda$(1710) resonance as the $\Lambda^*(70,{}^2 1,2,0^+)$ state, considering the huge width of the $\Lambda$(1710) resonance. The resonances $\Lambda(2070)$$\frac{3}{2}^{+}$ reported by BnGa group in 2019 \cite{Sara2019} is assigned as the $\Lambda^*(70,{}^2 1,2,2^+)$ state in Ref. \cite{Klempt2020}, but this work prefers it being the $\Lambda(70,{}^4 8,2,0^+)$ state. And the resonance $\Lambda(2110)$$\frac{5}{2}^{+}$
may be assigned as the $\Lambda(70,{}^4 8,2,2^+)$ state.

For the $L=1$ first excitation band, we may interpret the $\Lambda$(1800)$\frac{1}{2}^{-}$ and $\Lambda$(1830)$\frac{5}{2}^{-}$ resonances as three-quark states in the $\Lambda(70,{}^4 8,1,1^-)$ multiplet, as shown in table \ref{tab:posN}. The spin $\frac{3}{2}^{-}$ state, the spin-orbit coupling partner of the $\Lambda$(1800)$\frac{1}{2}^{-}$ and $\Lambda$(1830)$\frac{5}{2}^{-}$ resonances has not been reported. The two predicted flavor-singlet and two flavor-octet $\Lambda$ states of spin $\frac{1}{2}^{-}$ and $\frac{3}{2}^{-}$ lay rather far from the observed lowest negative-parity $\Lambda$ resonance states, thus we may not make any matching for these four negative-parity $\Lambda$ resonances in the pure 3q picture.

\begin{table}[b]
\caption{$q^3 s\bar q$ ground state pentaquark masses in ${\rm MeV}$. $J^P$ is the total angular momentum with $P$ being parity, and $S$ is the total spin of ground state pentaquarks, from the coupling of the spins of $q^3s$ and $\bar q$.}
\label{tab:dd3}
\begin{center}
\begin{indented}
	\item[]
\begin{tabular}{lccc}
\br
$J^P$ &
$q^3 s \bar q$ configurations &
$(S^{q^3s}, S^{\bar q}, S)$ &
 $M (q^3 s\bar q)$ \\
\hline
$\frac{5}{2}^{-}$  & $\Psi^{csf}_{[211]_{C}[31]_{FS}[31]_{F}[4]_{S}}(q^3s\bar q)$ & (2,1/2,5/2) &  2408
\\
$\frac{3}{2}^{-}$  & $\Psi^{csf}_{[211]_{C}[31]_{FS}[4]_{F}[31]_{S}}(q^3s\bar q)$ & (1,1/2,3/2) &  2392
\\
$\phantom{-}$ & $\left\lbrace \begin{array}{c} \Psi^{csf}_{[211]_{C}[31]_{FS}[31]_{F}[4]_{S}}(q^3 s \bar q) \\ \Psi^{csf}_{[211]_{C}[31]_{FS}[31]_{F}[31]_{S}}(q^3 s \bar q) \end{array} \right\rbrace$ & \makecell[c]{(2,1/2,3/2) \\ (1,1/2,3/2)} &  $\left( \begin{array}{c}1966  \\2407 \end{array} \right)$
\\
$\phantom{-}$ & $\Psi^{csf}_{[211]_{C}[31]_{FS}[211]_{F}[31]_{S}}(q^3s\bar q)$ & (1,1/2,3/2)   & 2116
\\
$\phantom{-}$ & $\Psi^{csf}_{[211]_{C}[31]_{FS}[22]_{F}[31]_{S}}(q^3s\bar q)$ & (1,1/2,3/2)  & 2229
\\
$\frac{1}{2}^{-}$ & $\Psi^{csf}_{[211]_{C}[31]_{FS}[4]_{F}[31]_{S}}(q^3s\bar q)$ & (1,1/2,1/2)  & 2659
\\
$\phantom{-}$ & $\left\lbrace \begin{array}{c}\Psi^{csf}_{[211]_{C}[31]_{FS}[31]_{F}[31]_{S}}(q^3 s \bar q)  \\ \Psi^{csf}_{[211]_{C}[31]_{FS}[31]_{F}[22]_{S}}(q^3 s \bar q) \end{array} \right\rbrace$  & \makecell[c]{(1,1/2,1/2)  \\(0,1/2,1/2)}   & $\left( \begin{array}{c}
2162  \\2314 \end{array} \right)$
\\
 $\phantom{-}$ & $\left\lbrace \begin{array}{c} \Psi^{csf}_{[211]_{C}[31]_{FS}[211]_{F}[31]_{S}}(q^3 s \bar q)  \\ \Psi^{csf}_{[211]_{C}[31]_{FS}[211]_{F}[22]_{S}}(q^3 s \bar q) \end{array} \right\rbrace$  & \makecell[c]{(1,1/2,1/2)  \\(0,1/2,1/2)}  & $\left( \begin{array}{c} 1742  \\2052\end{array} \right)$
\\
$\phantom{-}$ & $\Psi^{csf}_{[211]_{C}[31]_{FS}[22]_{F}[31]_{S}}(q^3s\bar q)$ & (1,1/2,1/2)  & 1894
\\
\br
\end{tabular}
\end{indented}
\end{center}
\end{table}

\section{Pentaquark components in negative-parity $\Lambda$ resonances}\label{sec:LPM}

As shown in the previous section, most $\Lambda$ resonances are reasonably described in the three-quark picture, except for the two pairs of the lowest negative-parity $\Lambda$ resonances. The flavor-singlet and flavor-octet $\Lambda$ states may be mixed to give the mass of $\Lambda$(1405) and other hyperon resonances, as shown in Refs. \cite{Klempt2020,Kiss2011,Mei2020}. In the present model, however, the masses of the $\Lambda_1$ singlet-flavor state and $\Lambda_8$ octet-flavor state are so close that one may make the mixture unreasonable for deriving the mass of the lowest negative-parity $\Lambda$ resonances. In this work, we assume that the $l=1$ negative-parity $\Lambda$ states may mix with the $q^3 s\bar q$ ground pentaquark states which are also of negative-parity.

The $q^3s\bar q$ pentaquarks are studied in the Hamiltonian in equation (\ref{eqn::ham}), where we have considered the couplings among different configurations due to the one-gloun-exchange (OGE) hyperfine interaction. Listed in table \ref{tab:dd3} are the theoretical masses of $q^3 s\bar q$ ground state pentaquarks. There are one $5/2$, five $3/2$ and six $1/2$ states in the spectra. As shown in Ref. \cite{Kai2019PRC}, the color and total spin-flavor parts of pentaquark states are always in the $[211]_{C}$ and $[31]_{SF}$ configurations. The $q^3s$ $[31]$ flavor configurations can be [4], [31] and [211], where the two possible Weyl tableaux $\young(uud,s)$ and $\young(uus,d)$ for the $q^3s$ $[31]$ flavor configuration correspond respectively to isospin $I=3/2$ and $I=1/2$ $q^3s$ components while the flavor configuration [4] leads to isospin $I=1$ and 2 pentaquark states. We focus only on isospin $I=0$ pentaquark states which may mix with negative-parity $q^2 s$ states.

Excluding the spin $\frac{5}{2}^{-}$ state, we have only four spin $\frac{3}{2}^{-}$ and five $\frac{1}{2}^{-}$ $q^3 s\bar q$ ground pentaquark states which may mix up with the two pairs of the low $p$-wave $q^2s$ states shown in the lower block of table \ref{tab:posN}. We mix the two $q^2s$ three-quark states and one $q^{3}s \bar q$ pentaquark state,
\begin{eqnarray}\label{threemw}
\left( \begin{array}{c} \psi_{1} \\ \psi_{2} \\ \psi_{3} \end{array} \right)=U
\left( \begin{array}{c} \Lambda_1 \\ \Lambda_8  \\ q^{3}s \bar q \end{array} \right) \nonumber \\
\end{eqnarray}
with U takes the form as,

{\small{
\begin{eqnarray*}
\hspace{-0.15\columnwidth}
\left(
\begin{array}{ccc}
 \cos (\theta ) \cos (\psi ) \cos (\phi )+\sin (\psi ) \sin (\phi ) & \cos (\psi ) \sin (\phi )-\cos (\theta ) \sin (\psi ) \cos (\phi ) & \sin (\theta ) \cos (\phi ) \\
 \sin (\psi ) \cos (\phi )-\cos (\theta ) \cos (\psi ) \sin (\phi ) & \cos (\theta ) \sin (\psi ) \sin (\phi )+\cos (\psi ) \cos (\phi ) & -\sin (\theta ) \sin (\phi ) \\
 -\sin (\theta ) \cos (\psi ) & \sin (\theta ) \sin (\psi ) & \cos (\theta ) \\
\end{array}
\right)
\end{eqnarray*}}}

\begin{table}[t]
\caption{The three-state mixtures of $q^2s$ singlet $\Lambda_1^*$, octet $\Lambda_8$ and ground $q^3s\,\bar q$ pentaquark states. The $\Lambda_1^*$ states, $\Lambda_8$ states for spin $1/2^-$ and $3/2^-$ take the theoretical values from table \ref{tab:posN}. All the possible $1/2^-$ and $3/2^-$ pentaquark states and masses (in MeV) are from table \ref{tab:dd3}.}
\label{threem3}
\small{
\begin{tabular}{@{}lccccccccc}
\br
 $\psi_1$ State &  $\psi_2$ State &  $\psi_3$ State & $J^{P}$ & $\Lambda_1$ State & $\Lambda_8$ State &  $\theta$ & $\phi$ & $\psi$ & $q^3 s\bar q$ states \\
\hline
1405 & 1670 ($\Lambda(1670)$) & 2248   & $\frac{1}{2} ^{-}$ & 1582 & 1579 & $157.3^{\circ}$ & $135.0^{\circ}$ & $90.4^{\circ}$ &  2162\\
1405 & 1670 ($\Lambda(1670)$) & 2400  & $\frac{1}{2} ^{-}$ & 1582 & 1579 & $159.9^{\circ}$ & $135.0^{\circ}$ & $90.3^{\circ}$ &  2314\\
1405 & 1670 ($\Lambda(1670)$) & 1828 ($\Lambda(1800)$) & $\frac{1}{2} ^{-}$ & 1582 & 1579 & $132.8^{\circ}$ & $133.0^{\circ}$ & $94.4^{\circ}$ &  1742\\
1405 & 1670 ($\Lambda(1670)$) & 2138 & $\frac{1}{2} ^{-}$ & 1582 & 1579 &  $154.6^{\circ}$ & $134.9^{\circ}$ & $90.6^{\circ}$ & 2052\\
1405 & 1670 ($\Lambda(1670)$) & 1980   & $\frac{1}{2} ^{-}$ & 1582 & 1579 &  $148.2^{\circ}$ & $134.9^{\circ}$ & $90.6^{\circ}$ & 1894\\
\mr
1510 & 1660 ($\Lambda(1690)$) & 1977 ($\Lambda(2050)$) & $\frac{3}{2} ^{-}$ & 1590 & 1591 & $169.3^{\circ}$ & $136.6^{\circ}$ & $88.0^{\circ}$  &  1966\\
1510 & 1660 ($\Lambda(1690)$) & 2418  & $\frac{3}{2} ^{-}$ & 1590 & 1591 & $173.1^{\circ}$ & $137.0^{\circ}$ & $87.6^{\circ}$  &  2407\\
1510 & 1660 ($\Lambda(1690)$) & 2127 & $\frac{3}{2} ^{-}$ & 1590 & 1591 & $171.2^{\circ}$ & $137.3^{\circ}$ & $87.3^{\circ}$  &  2116\\
1510 & 1660 ($\Lambda(1690)$) & 2240   & $\frac{3}{2} ^{-}$ & 1590 & 1591 &  $172.1^{\circ}$ & $137.6^{\circ}$ & $87.0^{\circ}$  & 2229\\
\br
\end{tabular}}
\end{table}

where $\psi_{1},\psi_{2},\psi_{3}$ are the resulting three negative-parity physical states from low to high energies corresponding to the observed negative-parity $\Lambda$ resonances, and $\Lambda_1$, $\Lambda_8$ and $q^{3}s \bar q$ stand for the flavor-singlet and flavor-octet $q^2s$ states in table \ref{tab:posN} and the ground $q^3s \bar q$ pentaquark states in table \ref{tab:dd3} respectively. The $3\times 3$ unitary matrix $U$ is parameterized as in Ref. \cite{Gil2018}, which is made of three rotation matrices of the angles $\theta$, $\psi$ and $\phi$.
The mixing angles are in the domain $0 \leq \theta \leq \pi$, $-\pi < \phi \leq \pi$ and $0 \leq \psi < \pi$. Through the unitary transformation of the mass matrices, one get the mass relation,
\begin{eqnarray}\label{meq}
M_{\Lambda_1}+ M_{\Lambda_8}+M_{q^3s \bar q}= M_{\psi_1}+M_{\psi_2}+M_{\psi_3}
\end{eqnarray}
where $M_{\Lambda_1}$, $M_{\Lambda_8}$ and $M_{q^3s \bar q}$ are the masses of the $\Lambda_1$ flavor-singlet state, the $\Lambda_8$ flavor-octet state and the ground $q^3s \bar q$ pentaquark state. $M_{\psi_3}$ and the mixing angles are determined by adjusting the $J=1/2$ lower states $\psi_1$ and $\psi_2$ respectively to $\Lambda(1405)$ and $\Lambda(1670)$, and the $J=3/2$ lower states $\psi_1$ and $\psi_2$ to $\Lambda(1520)$ and $\Lambda(1690)$. Listed in table \ref{threem3} are all possible three-state mixtures of the ground $q^3 s \bar q$ pentaquark states with $J^P=\frac{1}{2}^{-}$ and $\frac{3}{2}^{-}$ in table \ref{tab:dd3} with the $q^2s$ singlet $\Lambda_1$ and octet $\Lambda_8$ pairs. And the mixing matrices in equation (\ref{threemw}) show the composition of the physical states in terms of two $q^2s$ states and the lowest pentaquark state. Since it is impossible to derive $M_{\psi_1}=1520$ MeV and $M_{\psi_2}=1690$ MeV in the present model, we let 1510 MeV and 1660 MeV for $\Lambda(1520)$ and $\Lambda(1690)$ respectively.

\section{Discussion and Summary}\label{sec:SUM}
In this work, the masses of low-lying $q^2s$ states and ground $q^3s\bar q$ states are evaluated. Most positive-parity $\Lambda$ resonances and some negative-parity resonances are well reproduced in the three-quark picture. A tentative interpretation of the low-lying negative-parity $\Lambda$ resonances $\Lambda(1405)$, $\Lambda(1520)$, $\Lambda(1670)$ and $\Lambda(1690)$ is proposed, that is, the resonances might be the mixture of the $p$-wave $q^2s$ states and ground $q^3s\bar q$ pentaquark states. The theoretical results in equation (\ref{threemw}) have shown that $\Lambda(1405)1/2^-$ and $\Lambda(1520)3/2^-$ are mainly $q^2s$ states, but $\Lambda(1670)1/2^-$ and $\Lambda(1690)3/2^-$ contain considerable $q^3s \bar q$ pentaquark contributions. Except for the $\Lambda(1380)1/2^-$ resonance which can not be interpreted either as a three-quark or the mixture of three-quark and pentaquark states, we have reproduced the whole $\Lambda$ baryon spectrum for $N\leq2$ bands. 

As shown in table \ref{threem3}, the lowest $J=1/2$ and $J=3/2$ $M_{\psi_3}$ states are close to the $\Lambda(1800)$$1/2^-$ and $\Lambda(2050)\frac{3}{2} ^{-}$ resonance, respectively.
The 3-star $\Lambda(1800)$$1/2^-$ resonance of $M= 1800 \pm 50$ MeV and $\Gamma=200 \pm 50$ MeV is tentatively assigned in the $q^2s$ $\Lambda(70,{}^4 8,1,1^-)$ multiplet in Section \ref{sec:GTM}. However, the much larger BW width of the $\Lambda(1800)$$1/2^-$ resonance than the spin-orbit pair state $\Lambda(1830)\frac{5}{2}^{-}$ and the fact that Osaka group \cite{Kama2014,Kama2015} and JPAC \cite{Fern2016} analysis found no evidence of this state make it debatable that the $\Lambda(1800)$$1/2^-$ resonance is a pure three-quark state. One may make a bold guess that the $\Lambda(1800)\frac{1}{2}^{-}$ may be composed of two resonances, one is the $q^2s$ $\Lambda(70,{}^4 8,1,1^-)$ state and another the highest state of the three-state mixture of two lowest negative-parity $q^2s$ states and one ground $q^3s\bar q$ pentaquark state.

\begin{eqnarray}\label{threemw}
\left( \begin{array}{c} \Lambda(1405) \\ \Lambda(1670)  \\ \Lambda(1800) \end{array} \right)=\left(
\begin{array}{ccc}
 0.69 & -0.72 & 0.06 \\
 -0.52 & -0.44 & 0.73 \\
 -0.50 & -0.54 & -0.68 \\
\end{array}
\right)
\left( \begin{array}{c} \Lambda_1^* \\ \Lambda_8  \\ q^{3}s \bar q \end{array} \right)\nonumber \\
\left( \begin{array}{c} \Lambda(1520) \\ \Lambda(1690)  \\ \Lambda(2050) \end{array} \right)=\left(
\begin{array}{ccc}
 0.71 & -0.70 & -0.01 \\
 -0.69 & -0.70 & 0.19 \\
 -0.14 & -0.13 & -0.98 \\
\end{array}
\right)
\left( \begin{array}{c} \Lambda_1^* \\ \Lambda_8  \\ q^{3}s \bar q \end{array} \right)
\end{eqnarray}

The $\Lambda(2050)\frac{3}{2}^{-}$ resonance of $M=2056 \pm 22$ MeV and $\Gamma=493 \pm 61$ MeV was first reported by the KSU group \cite{Zhang22013}, and the bump was confirmed by JPAC analysis \cite{Fern2016} in 2016, with a smaller width $\Gamma=269 \pm 35$ MeV. The fact that the $P_c$(4450) state was turned out to be two much narrower $P_c$ states in the more precise LHCb experiment~\cite{LHCb3} has inspired one to guess that there may be more resonance states in the pole mass region of $\Lambda(2050)\frac{3}{2}^{-}$ since the width of this resonance is much larger than all other $\Lambda$ resonances in PDG.

%Two resonances $\Lambda(1800)\frac{1}{2}^{-}$ at M= $1800 \pm 50$ MeV with $\Sigma=200 \pm 50$ MeV and $\Lambda(2050)\frac{3}{2} ^{-}$ at M=$2056 \pm 22$ MeV with $\Sigma=493 \pm 61$ MeV are supposed to be the mixtures of the three-quark and pentaquark states.  so as the resonance $\Lambda(2050)\frac{3}{2}^{-}$. However,

\ack

 This research has received funding support from the NSRF via the Program Management Unit for Human Resources and Institutional Development, Research and Innovation (Grant No. B05F640055). Z.Z., A.K., T.Y. Htun, A.L., and Y.Y. all acknowledge support from Suranaree
University of Technology (SUT). A.K. would like to thank for the support of the Thailand Science Research and Innovation (TSRI) for their support through the Royal Golden Jubilee Ph.D. (RGJ-PHD) Program (Grant No. PHD/0242/2558) and the SUT-OROG scholarship (Contract No. 46/2558).

\appendix

\section{Explicit $q^2s$ $\Lambda$ wave functions}\label{ap:Lams}
In this Appendix the $q^2s$ color-orbital-spin-flavor wave functions with the principle quantum number $N\leq 2$ are listed in table \ref{tab:osf}, where $\chi_i$, $\Phi_j$, and $\phi^{N'}_{L'M'y}$ are the spin, flavor, and spatial wave functions, respectively. The $SU(3)_F$ decuplet states are excluded since $\Sigma$ and $\Sigma^*$ resonances are not discussed in the present work.

\begin{table}[t]
\caption{\label{tab:osf}Explicit $q^2s$ color-orbital-spin-flavor wave functions for isospin I=0. All the color wave function is $\psi^{c}_{[111]}$ singlet.}
\begin{center}{\small{
\begin{tabular}{@{}lllll}
\br
\multicolumn{1}{l}{\phantom{-}}&
   \multicolumn{1}{c}{$\phantom{-}$$SU(6)_{SF}$}&
   \multicolumn{1}{c}{$l^P$}&
   \multicolumn{2}{c}{$SU(6)_{SF}$$\times O(3)$ wave functions} \\
   \hline
   \multicolumn{1}{l}{$\phantom{-}N$} &
   \multicolumn{1}{c}{$\phantom{-}$Representations} &
   \multicolumn{1}{c}{O(3)}&
   \multicolumn{1}{l}{$\phantom{-}$$SU(3)_F$ singlet} &
   \multicolumn{1}{l}{$\phantom{-}$$SU(3)_F$ octet}\\[2pt]
\hline\\[-11pt]
   $\phantom{-}0$ & $\phantom{-}56$ & $0^{+}$ & $\phantom{-}$ & $J^{P}=\frac{1}{2}^{+}$\\[2pt]
    $\phantom{-}$ & $\phantom{-}$ & $\phantom{-}$ & $\phantom{-}$  & $\frac{1}{\sqrt{2}}\phi^0_{00s} (\Phi_{\lambda} \chi_{\lambda}+\Phi_{\rho} \chi_{\rho})$\\[2pt]
    \hline
   $\phantom{-}1$ & $\phantom{-}70$ & $1^{-}$  & $ \; J^{P}=\frac{1}{2}^{-}\; , \;\frac{3}{2}^{-}$ & $J^{P}=\frac{1}{2}^{-}\; , \;\frac{3}{2}^{-}$\\[2pt]
   $\phantom{-}$ & $\phantom{-}$ & $\phantom{-}$  & $\frac{1}{\sqrt{2}}(\chi_{\rho}\phi^1_{1m\lambda}- \chi_{\lambda}\phi^1_{1m\rho} )\Phi_{A} $   & $\frac{1}{2} [\phi^1_{1m\rho} (\Phi_{\lambda} \chi_{\rho}+\Phi_{\rho} \chi_{\lambda})+\phi^1_{1m\lambda} (\Phi_{\rho} \chi_{\rho}-\Phi_{\lambda} \chi_{\lambda})]$ \\[2pt]
   $\phantom{-}$ & $\phantom{-}$ & $\phantom{-}$  & $\phantom{-}$   & $J^{P}=\frac{1}{2}^{-}\; , \;\frac{3}{2}^{-},\;\frac{5}{2}^{-}$\\[2pt]
   $\phantom{-}$ & $\phantom{-}$ & $\phantom{-}$ &  $\phantom{-}$   & $\frac{1}{\sqrt{2}} \chi_{S} (\phi^1_{1m\lambda} \Phi_{\lambda}+\phi^1_{1m\rho} \Phi_{\rho})$\\[2pt]
   \hline
   $\phantom{-}2$ & $\phantom{-}56$ & $0^{+}$ & $\phantom{-}$ & $J^{P}=\frac{1}{2}^{+}$\\[2pt]
    $\phantom{-}$ & $\phantom{-}$ & $\phantom{-}$ & $\phantom{-}$   & $\frac{1}{\sqrt{2}}\phi^2_{00s} (\Phi_{\lambda} \chi_{\lambda}+\Phi_{\rho} \chi_{\rho})$ \\[2pt]
   \hline
   $\phantom{-}$ & $\phantom{-}70$ & $0^{+}$ & $ J^{P}=\frac{1}{2}^{+}$ & $J^{P}=\frac{1}{2}^{+}$\\[2pt]
   $\phantom{-}$ & $\phantom{-}$ & $\phantom{-}$ & $\frac{1}{\sqrt{2}}(\chi_{\rho}\phi^2_{00\lambda}- \chi_{\lambda}\phi^2_{00\rho} )\Phi_{A} $  & $\frac{1}{\sqrt{2}}[\phi^2_{00\rho} (\Phi_{\lambda} \chi_{\rho}+\Phi_{\rho} \chi_{\lambda})+\phi^2_{00\lambda} (\Phi_{\rho} \chi_{\rho}-\Phi_{\lambda} \chi_{\lambda})]$ \\[2pt]
   $\phantom{-}$ & $\phantom{-}$ & $\phantom{-}$ & $\phantom{-}$  & $J^{P}=\frac{3}{2}^{+}$\\[2pt]
   $\phantom{-}$ & $\phantom{-}$ & $\phantom{-}$ & $\phantom{-}$   & $\frac{1}{\sqrt{2}} \chi_{S} (\phi^2_{00\lambda} \Phi_{\lambda}+\phi^2_{00\rho} \Phi_{\rho})$\\[2pt]
   \hline
      $\phantom{-}2$ & $\phantom{-}20$ & $1^{+}$  & $J^{P}=\frac{1}{2}^{+}\; , \;\frac{3}{2}^{+}, \;\frac{5}{2}^{+}$ & $J^{P}=\frac{1}{2}^{+}\; , \;\frac{3}{2}^{+}$\\[2pt]
   $\phantom{-}$ & $\phantom{-}$ & $\phantom{-}$ &  $\phi^2_{1mA}  \chi_{S}\Phi_{A}$ & $\phi^2_{1mA} (\Phi_{\rho} \chi_{\rho}-\Phi_{\lambda} \chi_{\lambda})$\\[2pt]
   \hline
    $\phantom{-}2$ & $\phantom{-}56$ & $2^{+}$  & $\phantom{-}$ &$J^{P}=\frac{3}{2}^{+}\; , \;\frac{5}{2}^{+}$\\[2pt]
    $\phantom{-}$ & $\phantom{-}$ & $\phantom{-}$ & $\phantom{-}$   & $\frac{1}{\sqrt{2}} \phi^2_{2mS} (\Phi_{\rho} \chi_{\rho}+\Phi_{\lambda} \chi_{\lambda})$\\[2pt]
   \hline
   $\phantom{-}$ & $\phantom{-}70$ & $2^{+}$ & $ J^{P}=\frac{3}{2}^{+}\; , \;\frac{5}{2}^{+}$ & $J^{P}=\frac{3}{2}^{+}\; , \;\frac{5}{2}^{+}$\\[2pt]
   $\phantom{-}$ & $\phantom{-}$ & $\phantom{-}$ & $\frac{1}{\sqrt{2}}(\chi_{\rho}\phi^2_{2m\lambda}- \chi_{\lambda}\phi^2_{2m\rho} )\Phi_{A} $   & $\frac{1}{2} [\phi^2_{2m\rho} (\Phi_{\lambda} \chi_{\rho}+\Phi_{\rho} \chi_{\lambda})+\phi^2_{2m\lambda} (\Phi_{\rho} \chi_{\rho}-\Phi_{\lambda} \chi_{\lambda})]$\\[2pt]
   $\phantom{-}$ & $\phantom{-}$ & $\phantom{-}$ & $\phantom{-}$  & $J^{P}=\frac{1}{2}^{+}\; , \;\frac{3}{2}^{+} \;,\;\frac{5}{2}^{+} \;,\;\frac{7}{2}^{+}$ \\[2pt]
   $\phantom{-}$ & $\phantom{-}$ & $\phantom{-}$ &  $\phantom{-}$ & $\frac{1}{\sqrt{2}} \chi_{S} (\phi^2_{2m\lambda} \Phi_{\lambda}+\phi^2_{2m\rho} \Phi_{\rho})$ \\
  \br
\end{tabular}}}
\end{center}
\end{table}

The explicit flavor wave functions of flavor-singlet $\Lambda_1$ and flavor-octet $\Lambda_8$ are shown below:
\begin{eqnarray}
\Phi^{\Lambda_1}_{A} &=& \frac{1}{\sqrt{6}} (uds-dus+sud-usd+dsu-sdu), \nonumber\\
\Phi^{\Lambda_8}_{\lambda} &=& \frac{1}{2} (sud+usd-dsu-sdu), \nonumber\\
 \Phi^{\Lambda_8}_{\rho}&=& \frac{1}{\sqrt{12}} (2uds-2dus+sdu+usd-dsu-sud) \nonumber\\
\end{eqnarray}

\section*{References}
\bibliographystyle{unsrt}
%\bibliography{/home/thiago/bibtex/articles,/home/thiago/bibtex/books}

\begin{thebibliography}{}
\bibitem{Star2001}
A. Starostin, B. M. K. Nefkens, E. Berger, M. Clajus, A. Marusic, S. McDonald et al. (The Crystal Ball Collaboration), Measurement of $K^-\rightarrow{p} \eta \Lambda$ near threshold, Phys. Rev. C {\bf 64}, 055205 (2001).

% doi:10.1103/PhysRevD.34.2809
% DOI:https://doi.org/10.1103/PhysRevD.34.2809
\bibitem{Prakhov2004}
S. Prakhov, B. M. K. Nefkens, C. E. Allgower, V. Bekrenev, W. J. Briscoe, M. Clajus et al. (Crystal Ball Collaboration), Reaction ${K}^-p \rightarrow{\pi}^{0}{\pi}^{0}{\Lambda}$ from ${p}_{K^-}=514$ to 750 $MeV/c$, Phys. Rev. C {\bf 69}, 042202(R) (2004).

\bibitem{Prakhov2009}
S. Prakhov, B. M. K. Nefkens, V. Bekrenev, W. J. Briscoe, N. Knecht, A. Koulbardis et al. (Crystal Ball Collaboration), Measurement of ${\pi}^{0}{\Lambda}$, ${\overline{K}}^{0}n$, and ${\pi}^{0}\Sigma^{0}$ production in ${K}^-p$ interactions for ${p}_{K^-}$ between 514 and 750 MeV/$c$, Phys. Rev. C {\bf 80}, 025204 (2009).
%doi = {10.1103/PhysRevC.80.025204}

\bibitem{Zhang12013}
H. Zhang, J. Tulpan, M. Shrestha, and D. M. Manley, Partial-wave analysis of $\overline{K}N$ scattering reactions Phys. Rev. C. {\bf 88}, 035204 (2013).

\bibitem{Moriya2013}
K. Moriya et al. (CLAS Collaboration), Measurement of the ${\Sigma}{\pi}$ photoproduction line shapes near the ${\Lambda}(1405)$, Phys. Rev. C. {\bf 87},
035206 (2013).
%doi:10.1103/PhysRevC.87.035206
\bibitem{Moriya2014}
K. Moriya et al. (CLAS Collaboration), Spin and parity measurement of the $\mathrm{{\Lambda}}(1405)$ baryon, Phys. Rev. Lett. {\bf 112},
082004 (2014).
%doi:10.1103/PhysRevLett.112.082004

\bibitem{PDG2020}
P. A. Zyla et al. (Particle Data Group), Prog. Theor. Exp. Phys. {\bf 2020}, 083C01 (2020).

\bibitem{Zhang22013}
H. Zhang, J. Tulpan, M. Shrestha, and D. M. Manley, Multichannel parametrization of $\overline{K}N$ scattering amplitudes and extraction of resonance parameters, Phys. Rev. C. {\bf 88}, 035205 (2013).

\bibitem{Kama2014}
H. Kamano, S. X. Nakamura, T.-S. H. Lee, T. Sato, Dynamical coupled-channels model of ${K}^{-}p$ reactions: Determination of partial-wave amplitudes, Phys. Rev. C {\bf 90}, 065204 (2014).

\bibitem{Kama2015}
H. Kamano, S. X. Nakamura, T.-S. H. Lee, T. Sato, Dynamical coupled-channels model of ${K}^{-}p$ reactions. II. Extraction of ${\Lambda}^{*}$ and ${\Sigma}^{*}$ hyperon resonances, Phys. Rev. C. {\bf 92}, 025205 (2015); Erratum: [Phys. Rev. C {\bf 95}, 049903(E) (2017)]

\bibitem{Fern2016}
C. Fern\'andez-Ram\'{\i}rez, I. V. Danilkin, D. M. Manley, V. Mathieu, and A. P. Szczepaniak, Coupled-channel model for $\overline{K}N$ scattering in the resonant region, Phys. Rev. D. {\bf 93},
034029 (2016).

\bibitem{Matveev2019}
M. Matveev, A. V. Sarantsev, V. A. Nikonov, A. V. Anisovich, U. Thoma and E. Klempt, Hyperon I: Partial-wave amplitudes for $K^-p$ scattering, Eur. Phys. J. A {\bf 55}, 179 (2019).

\bibitem{Sara2019}
 A. V. Sarantsev, M. Matveev, V. A. Nikonov, A. V. Anisovich, U. Thoma and E. Klempt, Hyperon II: Properties of excited hyperons, Eur. Phys. J. A {\bf 55},180 (2019).

\bibitem{Aniso2020}
A. V. Anisovich, A. V. Sarantsev, V. A. Nikonov, V. Burkert, R. A. Schumacher, U. Thoma, and E. Klempt. Hyperon III: $K^{-} p-\pi\Sigma$ coupled-channel dynamics in the $\Lambda (1405)$ mass region, Eur. Phys. J. A {\bf 56}(5),139 (2020).

\bibitem{Klempt2020}
E. Klempt, V. Burkert, U. Thoma, L. Tiator and R. Workman, $\Lambda$
and $\Sigma$ Excitations and the Quark Model, Eur. Phys. J. A {\bf 56}, 261 (2020).
%
%\bibitem{Aniso2021}
%A. V. Anisovich, Excitation of hyperon resonances in photoproduction, Eur. Phys. J. A {\bf 57}, 236 (2021).

\bibitem{Bazzi2011}
M. Bazzi et al. (SIDDHARTA Collaboration), A new measurement of kaonic hydrogen X-rays, Phys. Lett. B {\bf 704}, 113 (2011).
%https://doi.org/10.1016/j.physletb.2011.09.011

\bibitem{Bazzi2012}
M. Bazzi et al. (SIDDHARTA Collaboration), Kaonic hydrogen X-ray measurement in SIDDHARTA, Nucl. Phys. A {\bf 881}, 88 (2012).

\bibitem{Isgur1978PRD}
N. Isgur, and G. Karl, P-wave baryons in the quark model, Phys. Rev. D. {\bf 18}, 4187 (1978).

\bibitem{Isgur1979PRD}
R. Koniuk, and N. Isgur, baryon decays in a quark model with chromodynamics, Phys. Rev. D. {\bf 21}, 1868 (1980).

\bibitem{Grome1983}
D. Grome, The Mysterious Spin-Orbit Interactions in Baryons, Nonlocal Forces and the p Wave Resonances, Z. Phys. C. {\bf 18}, 249 (1983).

\bibitem{Gloz1998}
L. Y. Glozman, W. Plessas, K. Varga, and R. F. Wagenbrunn, Unified description of light- and strange-baryon spectra, Phys. Rev. D {\bf 58}, 094030 (1998).
\bibitem{Loring2001}
U. L\"oring, K. Kretzschmar, B. C. Metsch, and H. R. Petry, The light-baryon spectrum in a relativistic quark model with instanton-induced quark forces, Eur. Phys. J. A {\bf 10}, 447 (2001).

\bibitem{Santo2015}
E. Santopinto and J. Ferretti, Strange and nonstrange baryon spectra in the relativistic interacting quark-diquark model with a G\"ursey and Radicati-inspired exchange interaction, Phys. Rev. C. {\bf 92}, 025202 (2015).
% doi = {10.1103/PhysRevC.92.025202},
\bibitem{Fau2015}
R. N. Faustov, V. O. Galkin, Strange baryon spectroscopy in the relativistic quark model, Phys. Rev. D {\bf 92}, 054005 (2015).
%three body three three quark mixture

\bibitem{Capstick2000}
 S. Capstick and W. Roberts, Quark models of baryon masses and decays, Prog. Part. Nucl. Phys. {\bf 45},
S241 (2000), and references therein.

\bibitem{Alston1961}
M. H. Alston, L. W. Alvarez, P. Eberhard, M. L. Good, W. Graziano, H. K. Ticho, S. G. Wojcicki, Study of Resonances of the ${\Sigma}{-}{\pi}$ System, Phys. Rev. Lett. {\bf 6}, 698 (1961).

\bibitem{Dalitz1960}
R. H. Dalitz and S. F. Tuan, The phenomenological representation of $\bar K$-nucleon scattering and reaction amplitudes, Annals Phys. {\bf 10}, 307 (1960).

\bibitem{Dalitz1967}
R. H. Dalitz, T. C. Wong and G. Rajasekaran, Model Calculation for the $Y_0^*(1405)$ Resonance State, Phys. Rev. {\bf 153}, 1617 (1967)

\bibitem{Oller2001}
J. A. Oller and U.-G. Mei\ss ner, Chiral dynamics in the presence of bound states: kaon-nucleon interactions revisited, Phys. Lett. B {\bf 500}, 263 (2001).

\bibitem{Oller2003}
D. Jido, J. A. Oller, E. Oset, A. Ramos and U.-G. Mei\ss ner, Chiral dynamics of the two $\Lambda$(1405) states, Nucl. Phys. A {\bf 725}, 181 (2003).

\bibitem{Roca2013}
L. Roca and E. Oset, $\Lambda$(1405) poles obtained from $\pi^0 \Sigma^0$ photoproduction data, Phys. Rev. C {\bf 87}, 055201 (2013). 

\bibitem{Oset2013}
 L. Roca and E. Oset, Isospin 0 and 1 resonances from $\pi \Sigma$ photoproduction data, Phys. Rev. C {\bf 88}, 055206 (2013). 

\bibitem{Mai2015}
M. Mai and U.-G. Mei\ss ner, Constraints on the chiral unitary $\bar KN$ amplitude from $\pi \Sigma K^+$ photoproduction data, Eur. Phys. J. A {\bf 51}(3), 30 (2015). 

\bibitem{kaiser1997}
N. Kaiser, T. Waas and W. Weise, SU(3) Chiral Dynamics with Coupled Channels: Eta and Kaon Photoproduction, Nucl. Phys. A {\bf 612}, 297-320 (1997).

\bibitem{Oset2002}
E. Oset, A. Ramos and C. Bennhold, Low lying $S = -1$ excited baryons and chiral symmetry, Phys. Lett. B {\bf 527}, 99 (2002) Erratum: [Phys. Lett. B {\bf 530}, 260 (2002)].

\bibitem{Cieply2010}
A. Ciepl\'y and J. Smejkal, Separable potential model for $K^-N$ interactions at low energies, Eur. Phys. J. A {\bf 43}, 191 (2010).

\bibitem{Cieply2012}
 A. Ciepl\'y and J. Smejkal, Chirally motivated $\bar K$ N amplitudes for in-medium applications, Nucl. Phys. A {\bf 881}, 115 (2012).
 
 \bibitem{Ikeda2011}
Y. Ikeda, T. Hyodo and W. Weise, Improved constraints on chiral $SU(3)$ dynamics from kaonic hydrogen, Phys. Lett. B {\bf 706}, 63 (2011). 

\bibitem{Ikeda2011two}
Y. Ikeda, T. Hyodo and W. Weise, Chiral SU(3) theory of antikaon-nucleon interactions with improved threshold constraints, Nucl. Phys. A {\bf 881}, 98 (2012).

\bibitem{Guo2013}
Z. H. Guo and J. A. Oller, Meson-baryon reactions with strangeness -1 within a chiral framework, Phys. Rev. C {\bf 87}, 035202 (2013).

\bibitem{Mai2013}
M. Mai and U.-G. Mei\ss ner, New insights into antikaon-nucleon scattering and the structure of the $\Lambda$(1405), Nucl. Phys. A {\bf 900}, 51-64 (2013). 

\bibitem{Miyahara2018}
K. Miyahara, T. Hyodo and W. Weise, Construction of a local $\bar KN-\pi \Sigma-\pi \Lambda$ potential and composition of the $\Lambda$(1405), Phys. Rev. C {\bf 98}, 025201 (2018). 
 
 \bibitem{Cieply2016}
 A. Ciepl\'y, M. Mai, U.-G. Mei\ss ner and J. Smejkal, On the pole content of coupled channels chiral approaches used for the $\bar KN$ system, Nucl. Phys. A {\bf 954}, 17 (2016). 

\bibitem{Ren2021}
X. L. Ren, E. Epelbaum, J. Gegelia and U. G. Mei\ss ner, The $\Lambda$(1405) in resummed chiral effective field theory, Eur. Phys. J. C {\bf 81}: 582 (2021).

\bibitem{Wang2021}
Z. Y. Wang, H. A. Ahmed and C. W. Xiao, Is two-pole's $\Lambda$(1405) one state or two?, Eur. Phys. J. C {\bf 81}, 833 (2021)
%https://doi.org/10.1140/epjc/s10052-021-09633-4

\bibitem{Bruns2022}
 P. C. Bruns and A. Ciepl\'y, SU(3) flavor symmetry considerations for the KN coupled channels system, Nucl. Phys. A {\bf 1019}, 122378 (2022).
 
 \bibitem{Nakamura2008}
T. Nakamura, J. Sugiyama, N. Ishii, T. Nishikawa and M. Oka, Exotic quark structure of $\Lambda$(1405) in QCD sum rule, Phys. Lett. B {\bf 662}, 132 (2008).
 
\bibitem{Azi2018}
K. Azizi, B. Barsbay, and H. Sundu, Mass and residue of $\Lambda(1405)$ as hybrid and excited ordinary baryon, Eur. Phys. J. Plus {\bf 133}, 121 (2018).
%https://doi.org/10.1140/epjp/i2018-11928-9

\bibitem{Kiss2011}
L. S. Kisslinger, E. M. Henley, QCD: The $\Lambda(1405)$ as a hybrid, Eur. Phys. J. A {\bf 47}, 8 (2011).

\bibitem{Arima1994}
M. Arima, S. Matsui and K. Shimizu, $\Lambda(1405)$ and meson-baryon interaction in a quark model, Phys. Rev. C. {\bf 49}, 2831 (1994).

\bibitem{Sachi2007}
S. Takeuchi and K. Shimizu, $\Lambda(1405)$ as a resonance in the baryon-meson scattering coupled to the ${q}^{3}$ state in a quark model, Phys. Rev. C. {\bf 76}, 035204 (2007).

\bibitem{Hall2015}
J. M. M. Hall, W. Kamleh, D. B. Leinweber, B. J. Menadue, B. J. Owen, A. W. Thomas, and R. D. Young, Lattice QCD Evidence that the $\Lambda(1405)$ Resonance is an Antikaon-Nucleon Molecule, Phys. Rev. Lett. {\bf 114},
132002 (2015).
%10.1103/PhysRevLett.114.132002
\bibitem{Hall2017}
J. M. M. Hall, W. Kamleh, D. B. Leinweber, B. J. Menadue,
B. J. Owen, and A. W. Thomas, Light-quark contributions to the magnetic form factor of the $\Lambda(1405)$, Phys. Rev. D. {\bf 95}, 054510 (2017).


\bibitem{Kai2019PRC}
K. Xu, A. Kaewsnod, X. Y. Liu, S. Srisuphaphon, A. Limphirat, and Y. Yan, Complete basis for pentaquark wave function in group theory approach, Phys. Rev. C {\bf 100}, 065207 (2019).

\bibitem{Kai2020PRD}
K. Xu, A. Kaewsnod, Z. Zhao, X. Y. Liu, S. Srisuphaphon, A. Limphirat, and Y. Yan, Pentaquark components in low-lying baryon resonances, Phys. Rev. D {\bf 101}, 076025 (2020).


\bibitem{Mei2020}
U. G. Mei\ss ner, Two-Pole Structures in QCD: Facts, Not Fantasy!, Symmetry {\bf 12}, 981 (2020).


%https://doi.org/10.1140/epja/i2011-11008-5


%doi = {10.1103/PhysRevD.95.054510},


\bibitem{Gil2018}
J. J. Gil, Parametrization of $3\times 3$ unitary matrices based on polarization algebra, Eur. Phys. J. Plus {\bf 133}, 206 (2018).
%https://doi.org/10.1140/epjp/i2018-12032-0


\bibitem{LHCb3}
R. Aaij et al. (LHCb Collaboration), Observation of a Narrow Pentaquark State, $P_c(4312)^{+}$ , and of the Two-Peak Structure of the $P_c(4450)^{+}$, Phys. Rev. Lett. {\bf 122}, 222001 (2019).

\end{thebibliography}

\end{document}